\documentclass[10pt,twocolumn]{article}
\usepackage{natbib}
\usepackage{mybib}
 \usepackage{graphicx}
\begin{document}\title{A common trend in the chemical evolution 
of \\ Local Group dwarf spheroidals and Damped Ly\,$\alpha$ systems}\author{G. Vladilo$^1$, L. Sbordone$^2$, P. Bonifacio$^1$\\
(1) Osservatorio Astronomico di Trieste - INAF, Italy\\
(2) European Southern Observatory, Santiago, Chile}\date{}\maketitle 
\abstract{  
We compare chemical abundances 
of Local Group  dwarf spheroidals,
obtained from recent UVES/VLT observations,
and of high redshift Damped Lyman $\alpha$  systems (DLAs), corrected for dust effects. 
We focus, in particular, on the abundance ratio between
$\alpha$-capture elements and iron, $\alpha$/Fe, a well known
indicator of chemical evolution. 
Comparison of the data in the plane $\alpha$/Fe versus Fe/H   shows
a remarkable similarity between the dwarf spheroidals
and the DLAs, suggestive of   a common trend in their chemical evolution.
At any given  metallicity these two distinct types of astronomical targets
 show $\alpha$/Fe ratios
systematically lower than those of Milky Way stars. 
In terms  
 of chemical evolution models, this   suggests that, on average,   
dSph galaxies and DLA systems are characterized by lower,  
or more episodic, star formation rates than the Milky Way. 
 }

\section{Introduction}

With the advent of 10-m class optical telescopes equipped
with high resolution spectrographs,  Local Group dwarf
galaxies may now be observed on a star by star basis and abundance ratios
measured with sufficent  accuracy  for probing their chemical evolution
(see e.g. Shetrone et al. 2001). 
With the same type of instrumentation,
detailed abundances can be obtained at high redshift for the class
of quasar absorbers with highest HI column density, the Damped Lyman alpha (DLA)
systems (see e.g. Prochaska et al. 2001). 
Evidence is accumulating that these absorbers originate in the
interstellar gas of  high redshift galaxies. Study of their abundances can be
used to compare their chemical evolution properties with those of local dwarfs.
In the hierarchical bottom-up scenario for galaxy formation, dwarf galaxies
are the basic building blocks out of which larger galaxies are assembled.
If this picture is correct, one expects to find 
a significant fraction of dwarf galaxies  when looking at the high
redshift Universe. 
Comparing the chemical abundances of local dwarfs   with those of
high redshift DLAs may provide
 clues in favour or against this scenario. 
In this contribution we compare
abundances in Local Group dwarf spheroidals, described in the next
section, with those DLA systems, described in Section 3. 

\section{LG dwarf spheroidals}
 
We obtained abundances of  Local Group galaxies
using the UVES spectrograph fed by the Kueyen-VLT 8.2m
ESO telescope. 
For two giants in the Sgr dSph galaxy 
we determined the detailed
abundances  using
the UVES commissioning data (Bonifacio et al. 2000). 
Recently we have
increased the sample by analysing ten more giants
 Bonifacio et al. (2003).
The spectra have a resolution $\Delta\lambda/\lambda \simeq
43000$ and cover the range 480--680 nm.
 Each star was observed repeatedly
and the final S/N ratio on the coadded spectrum was in the range
20--40 at 530\,nm.
Some examples of spectra are shown in Fig. 1.
Effective temperatures were derived from the $(V-I)_0$ 
colour of Marconi et al. (1998),
through the calibration of Alonso et al. (1999),
adopting a mean reddening $E(V-I)=0.22$, $A_V = 0.55$
and a distance
modulus $m-M=16.95$  (Marconi et al. 1998).
The abundances were derived mainly from equivalent widths
using the WIDTH code (Kurucz 1993), although some lines were treated
with spectrum synthesis SYNTHE code (Kurucz 1993).
In both cases  {\em ad hoc } model atmospheres were computed with the
ATLAS code.
The surface gravities were adjusted to satisfy the iron ionization
equilibrium, and were all around log g = 2.5, as expected from the
location of the stars in the colour-magnitude diagram and the comparison
with theoretical isochrones.

\begin{figure}   \centering
 \includegraphics[clip=true,width=7cm,angle=0]{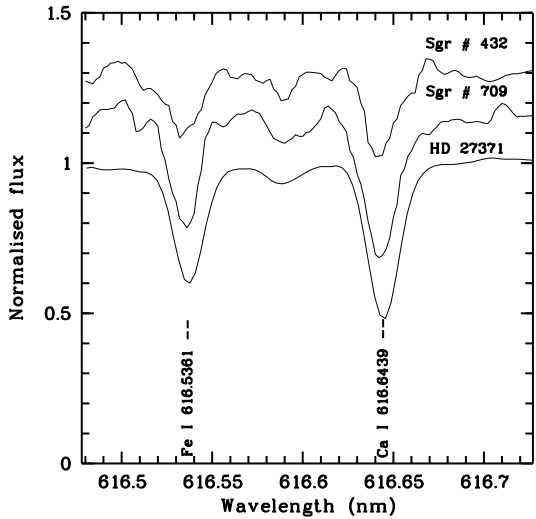} 
\caption{   Spectral region around
the Fe I 616.5 nm and Ca I 616.6 nm lines for two stars of the Sgr
dSph galaxy %
and for a comparison giant star of the Hyades.% 
} \label{fig1}\end{figure}

Two main results can be derived from our sample 
  of 12 giant stars:
(1)
 the Sgr dSph galaxy hosts a relatively metal-rich population,
       which in fact dominates our sample;
(2) the $\alpha$/Fe ratio  
       is solar or sub-solar, even for the most metal-poor star
       of the sample, star \# 432 with [Fe/H]=--0.83 and
       [$\alpha$/Fe]$\sim -0.1$. This chemical pattern is clearly distinct from what
observed in Milky Way stars of comparable metallicity.
Both results can be seen in Fig. 2, where we plot 
the ratio [$\alpha$/Fe] $\equiv$  0.5$\times$([Mg/Fe]+[Ca/Fe])
versus [Fe/H] for
the twelve stars of our
sample (solid circles) and for Milky Way stars (crosses). 
In Fig. 2 we also plot the same ratios 
measured in stars of other 7 Local Group dwarf spheroidals
(open circles), collected from the literature
(Shetrone et al. 2001, 2003).
All together, the data of the 8 dwarf spheroidals show a remarkable
continuity
in the  [$\alpha$/Fe] versus [Fe/H] plane. 
\section{ Damped Ly  $\alpha$ systems}

The most widely measured $\alpha$/Fe ratio in DLA systems
is Si/Fe, since both Si and Fe have several unsaturated transitions 
redshifted into the optical range (e.g. Prochaska et al. 2001). The Si/Fe ratios in DLAs
 can be significantly enhanced due to   dust depletion
and their interpretation  in
terms of chemical evolution requires a preliminary correction from dust
  effects  (Vladilo 1998).

 Recently we have presented a
new method aimed at correcting interstellar abundances for
dust depletion, taking into account possible variations of the chemical
composition of the dust (Vladilo 2002). The method can be applied
to interstellar measurements in galaxies with non solar composition
and, in particular, to DLA systems. The method  
requires  a determination of the column densities
of Fe, Zn, and the element X for which the ratio X/Fe
must be corrected. A guess of the intrinsic Zn/Fe ratio in the system
is also required.   
The method has been tested in the SMC, where the interstellar 
 ([Si/Fe],[Fe/H]) data
show a large scatter (Welty et al. 2001) and  differ from the typical
 stellar SMC values (Russell \& Dopita 1992)
 as a consequence of 
dust depletion, which is known to affect the interstellar measurements. 
As shown in Fig. 3, 
the application of the  method to the interstellar measurements
brings together all the interstellar and stellar data of the SMC.
 This result indicates the importance  
and the validity of the dust correction method.

 In Fig. 2 we plot the ([Si/Fe], [Fe/H]) data in DLA systems
corrected for dust 
  (triangles). The original measurements are 
based on  spectra collected with
the Keck and the VLT telescopes (see references in
Vladilo 2002). The dust corrected data in Fig. 2 have been obtained
assuming that the intrinsic Zn/Fe in DLAs is slightly oversolar, [Zn/Fe]=+0.1 dex,
as observed in Milky Way stars (Gratton et al. 2003). 
Should the intrinsic Zn/Fe ratio be solar, the corrected [Si/Fe] ratios would
be lower by $\simeq 0.1$ dex.

From Fig. 2 one can see  that
the ([$\alpha$/Fe], [Fe/H]) data of DLAs   follow the  
same trend shown by dwarf spheroidals of the Local Group.

 \begin{figure}%[htbp]   \centering
  \includegraphics[clip=true,width=7.2cm,angle=0]{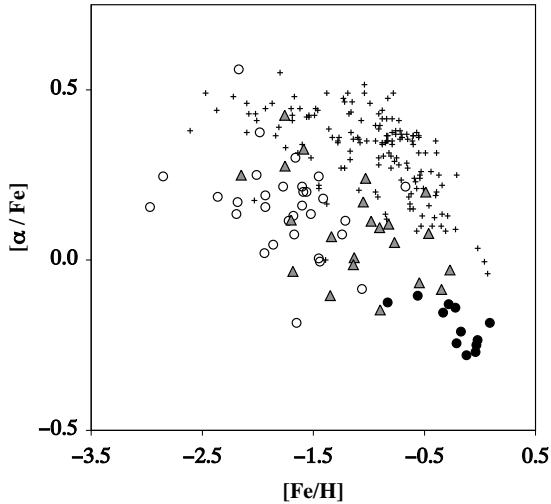} 
\caption{  Abundance ratios [$\alpha$/Fe] vs. [Fe/H]  
in stars of 8 dSph galaxies of the Local Group  (solid circles: Bonifacio et al. 2003;
open circles: Shetrone et al. 2001, 2003),
in DLA systems (triangles: Vladilo 2002), and in Milky Way stars (crosses: Gratton et al. 2003).
For the stellar measurements we adopt  
[$\alpha$/Fe] $\equiv$  0.5$\times$([Mg/Fe]+[Ca/Fe]); 
for DLAs [$\alpha$/Fe] $\equiv$ [Si/Fe], with dust corrected  abundances.
} \label{fig2}\end{figure}

 \begin{figure}%[htbp]  \centering
   \includegraphics[clip=true,width=7.5cm,angle=0]{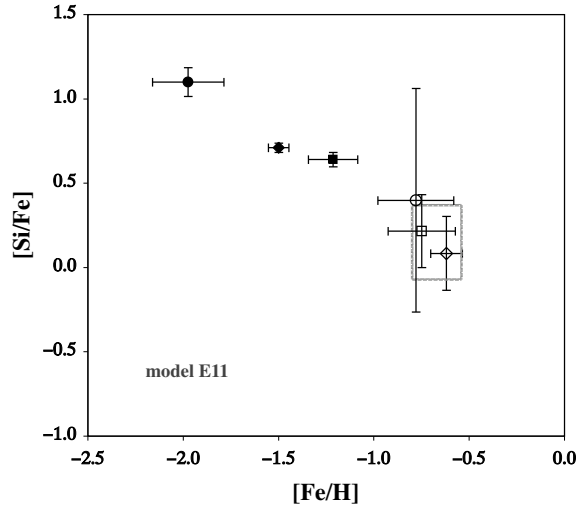} 
\caption{  Interstellar [Si/Fe] versus [Fe/H]
abundances in three SMC lines of sight. Filled symbols: measurements from Welty et al. (2001). Empty symbols: same
data corrected for dust depletion using the set of parameters E11
 (Vladilo 2002).
Box: range of SMC stellar measurements (Russell \& Dopita 1992).
} \label{fig3}\end{figure}

 \section{Discussion}

The distribution of the ([$\alpha$/Fe],[Fe/H]) data points in Fig. 2 has
several implications concerning the chemical evolution properties of
dSph galaxies of the Local Group   and DLA systems. 
 
 The remarkable continuity between our data points of the Sgr dSph (solid circles in Fig. 2)
 and the data of the other dSph galaxies collected from the literature (open circles)
 suggests that dwarf spheroidals lie on a similar evolutionary path, but some
 are more evolved than the others. 
 
 The similarity between local dwarf spheroidals (circles)
 and high redshift DLA systems (triangles) indicate that these two
 distinct types of objects
 share a common evolutionary path. This result is roubust against
 variations of   the input parameters in the dust correction procedure. 
 
 An origin of low redshift ($z \leq 1$) DLA systems in dwarf, compact and low surface brightness
 galaxies, with little contribution from spirals, was indicated by imaging studies 
 (Le Brun et al. 1997; Rao \& Turnshek 1998). The present analysis of abundance ratios
 indicates  that also at high redshift ($z \geq 1.5$) most
 DLA systems behave similarly to local dwarfs as far as the chemical evolution is concerned. 
 The [$\alpha$/Fe] values tend to decrease with metallicity for each class of objects
 considered in Fig. 2 (dSph stars, Milky Way stars, DLA systems).
 Chemical evolution models can explain such a decrease in terms of a time delay between an
 early enrichment due to type II SNae, rich in $\alpha$-capture elements, and a subsequent enrichment due
 to type Ia SNae, rich in Fe-peak elements (see e.g. Calura et al. 2003). 
 The decrease of  [$\alpha$/Fe] is predicted to be faster 
 for galaxies with high rates of star formation and slower for galaxies
 with low rates, or episodic star formation. 
 In this context the lower [$\alpha$/Fe] values of dSph stars and DLA systems,
 compared to those of Milky Way stars at the same level of  metallicity,
 may reflect a difference in the star formation rate, the Milky Way  
 having been characterized by a higher or more continuous star formation rate  
 than  dSph galaxies and DLA systems. 
\section{Conclusions}
 
We compared $\alpha$/Fe abundance ratios measured in
stars of  Local Group dSph galaxies, in stars of the Milky Way, 
and in DLA systems. 
 The  $\alpha$/Fe ratios of the Local Group include our recent measurements
of 12 stars in the Sgr dSph galaxy. By adding up literature data, a total
of 8 dwarf spheroidal galaxies were considered.  
 The $\alpha$/Fe ratios of the DLA systems were corrected for dust depletion
with a method recently developed by us and tested in the ISM of the SMC. 
The sample includes 22 DLA systems with Si/Fe data. 
 The analysis of the data in the plane $\alpha$/Fe versus Fe/H   shows
a remarkable similarity between the Local Group dwarf spheroidals
and the high-redshift DLA systems.  This result indicates that these
two distinct types of galaxies share a common trend of chemical evolution.
 
The similarity between high-redshift DLA systems ($z \geq 1.5$) and local dwarfs
derived from the abundance analysis is
in line with the results of imaging studies of low-redshift DLA systems ($z \leq 1$)
which indicate that most DLA galaxies are dwarf, compact, or low surface brightness
galaxies, with little contribution from spirals.
 
dSph stars and DLA systems have $\alpha$/Fe ratios
systematically lower than those of Milky Way stars
of same metallicity. 
In terms of the time delay interpretation
 of chemical evolution models, this result suggests that, on average, the 
dSph galaxies and DLA systems are characterized by lower  
(or more episodic) star formation rates than the Milky Way. 
This suggests that, with possible exceptions, DLAs do not represent
the young stage of Milky-Way type spirals. In addition, it is unlikely that 
DLAs could be major building blocks out of which the proto-Milky Way
was formed, considering that the vast majority of Milky-Way stars show significantly 
enhanced
$\alpha$/Fe ratios at low metallicity. 
 
%
%==============================================================================
% Acknowledgements (empty)
%===============================================================================
% References
%===============================================================================

\bibliographystyle{apj}

 \end{document}